\documentstyle[12pt]{article}
\def\pictures{y }
\ifx\pansw\pictures
\input prepictex
\input pictex
\input postpictex
\else
\fi
\newdimen\tdim
\newcommand{\mysection}[1]{\setcounter{equation}{0}\section{#1}}

\begin{document}
\newcommand{\nc}{\newcommand}
\nc{\beq}{\begin{equation}}     \nc{\eeq}{\end{equation}}
\nc{\beqa}{\begin{eqnarray}}    \nc{\eeqa}{\end{eqnarray}}
\nc{\lsim}{\begin{array}{c}\,\sim\vspace{-21pt}\\< \end{array}}
\nc{\gsim}{\begin{array}{c}\sim\vspace{-21pt}\\> \end{array}}
\nc{\create}{\hat{a}^\dagger}   \nc{\destroy}{\hat{a}}
\nc{\kvec}{\vec{k}}             \nc{\kvecp}{\vec{k}^\prime}
\nc{\kvecpp}{\vec{k}^{\prime\prime} }   \nc{\kb}{\bf k}
\nc{\kbp}{{\bf k}^\prime}       \nc{\kbpp}{{\bf k}^{\prime\prime} }
\nc{\bfk}{{\bf k}}              \nc{\cohak}{a_{{\bf k}}}
\nc{\cohap}{a_{{\bf p}}}        \nc{\cohaq}{a_{{\bf q}}}
\nc{\cohbk}{b^*_{{\bf k}}}      \nc{\cohbp}{b^*_{{\bf p}}}
\nc{\cohbq}{b^*_{{\bf q}}}      \nc{\Cop}{\hat{C}}
\nc{\ir}{\vert\, i\,\rangle}    \nc{\fr}{\vert\, f\,\rangle}
\nc{\il}{\langle\, i\,\vert}    \nc{\fl}{\langle\, f\,\vert}
\nc{\A}{A_{\mu} (x)}            \nc{\ivr}{\vert\, 0\,\rangle^{\rm\, in}}
\nc{\ivl}{^{\rm\, in}\langle\, 0\,\vert}
\nc{\ovr}{\vert\, 0\,\rangle^{\rm out}}
\nc{\ovl}{^{\rm out}\langle\, 0\,\vert}  \nc{\I}{{\rm in}}
\nc{\out}{{\rm out}}            \nc{\PA}{{\cal P} (A)}
\nc{\SA}{S^{\mu}_{\epsilon} }
\nc{\JPS}{J^\mu\left( x\,\vert\,\epsilon\right)}
\nc{\ffdool}{{\rm Tr}\, F\,\widetilde{F}}  \nc{\BF}{\{\A ,\Phi (x)\}}
\vskip .5in
\begin{titlepage}
\begin{center}
{\hbox to\hsize{October 1994 \hfill JHU-TIPAC-940018}}
{\hbox to\hsize{hep-ph/9410408   \hfill HUTP-A0/039}}
\vskip 1 in
{\Large \bf Anomalous Violation of Conservation Laws}
\\[0.125in]
{\Large \bf in Minkowski Space:} \\[0.125in]
{\Large \bf Spontaneously Broken Gauge Theories} \\
\vskip 0.5in

\begin{tabular}{cc}
\begin{tabular}{c}
{\bf Thomas M. Gould\footnotemark[1]}\\[.05in]
{\it Department of Physics and Astronomy}\\
{\it The Johns Hopkins University}\\
{\it Baltimore MD 21218 USA}\\[.15in]
\end{tabular}
\begin{tabular}{c}
{\bf Stephen D.H. Hsu}\footnotemark[2]\\[.05in]
{\it Lyman Laboratory of Physics} \\
{\it Harvard University} \\
{\it Cambridge  MA 02138 USA} \\[.15in]
\end{tabular}
\end{tabular}

\vskip .5 in
{\bf Abstract}\\[-0.05in]
\smallskip
\begin{quote}

We extend our previous results on the evolution of
quantum fermi fields in Minkowski gauge field backgrounds
to the case of spontaneously broken gauge theories.
We obtain a selection rule which relates the amount of
fermion number violation to the change in Higgs
winding number of the configuration. This selection rule
is applicable to any classical solution which
dissipates at both early and late times.
\end{quote}
\end{center}
\footnotetext[1]{gould@fermi.pha.jhu.edu}
\footnotetext[2]{hsu@hsunext.harvard.edu;
Address after January 1995:
Yale University, Sloan Physics Laboratory,

\hspace{0.2cm}Department of Physics, New Haven CT 06511.}
\end{titlepage}

\renewcommand{\thepage}{\arabic{page}}
\setcounter{page}{1}
\mysection{\, Introduction }

In previous work, we explored the phenomena of anomalous
fermion number violation in non-abelian gauge theories~\cite{us},
a subject studied several years ago by Christ~\cite{NC}.
We derived selection rules for fermion scattering amplitudes
which relate the violation of fermionic charge to the
change in winding between early and late times of
the vacuum part of the gauge field.
In this sequel, we will generalize our analysis to the
case of spontaneously broken gauge theories. We will again
find that the change in Chern--Simons number of the background field,
or its topological charge,  does not by itself determine
the amount of fermion number violation.

For any configuration which approaches a vacuum up to
additional ``radiation'' which dissipates sufficiently rapidly,
the amount of fermion number violation is determined solely by the
change in Chern--Simons number of the relevant {\it vacua} that the
configuration approaches at early and late times.
In particular,
our new selection rules will be applicable to any classical
solution of the field equations which approaches a vacuum in both
the far past and future.

We will use the notation and formalism of \cite{us},
with the additional complication that the background fields now consist
of gauge {\em and} scalar fields, $\BF$.
Our analysis should be valid for any model in which the gauge symmetry
is completely broken by vacuum expectation values of the scalar field(s).
For simplicity,
we will discuss a specific model of an $SU(2)$ gauge field
coupled to a complex scalar doublet.

The outline of our paper is as follows.
In section 2,
we define the various quantities with which we will be working,
in particular the operator formalism we will use to count fermions.
in the unitary gauge.
In section 3,
we derive our selection rule and discuss its implications.
In section 4,
we discuss the asymptotic behavior of exact solutions in the unitary gauge.
In section 5, we give our conclusions.

\mysection{Specifying the Problem}

Our purpose in this section is to give precise meaning
to the various mathematical quantities we will be using in this paper.
Further details and discussion can be found in \cite{us}.
We work in the context of quantum fermi fields $\Psi (x)$
in a finite volume of spatial size $L$.
The volume will ulitmately  be taken to infinity in a way explained
at the end of this section.

We consider $SU(2)$ doublet fermions with chiral couplings to
the $SU(2)$ gauge fields $\A $ and Yukawa couplings to the Higgs
doublet $\Phi (x)$.
These are the couplings of fermions in the electroweak theory,
neglecting the weak mixing angle.
The $U(1)$ fermion number (baryon plus lepton number)
current in this theory has an anomalous divergence, given by
\beq
\label{anom}
\partial_{\mu} J^{\mu} ~=~  \frac{g^2N}{16\pi^2}~ \ffdool ~,
\eeq
where $N$ is the number of fermion doublets.
The relation (\ref{anom}) is an operator equation,
and allows the replacement of the operator $\partial\!\cdot\! J$ with
the c-number on the right hand side of (\ref{anom}) in any matrix element.
The current is defined by
\beq
\label{JJ1}
J^{\mu} ~\equiv~
\lim_{\epsilon \rightarrow 0}\, J^\mu\left(x\,\vert\,\epsilon\right)  ~+~
\left\{ c.t. \right\} ~,
\eeq
where $\left\{ c.t.\right\}$ is a counterterm discussed below
and the gauge invariant, point split current is given by
\beq
\label{ps}
J^\mu\left(x\,\vert\,\epsilon\right) ~\equiv~
\overline{\Psi} (x + \epsilon/2)\,\gamma^\mu\, Q\,
{\cal P}\exp{\left(ig\int_{x - \epsilon/2}^{x + \epsilon/2} dy
\cdot A (y) \right)}
\,\Psi (x - \epsilon/2) ~.
\eeq
Here $A_\mu = T^a A^a_\mu$ and $T^a$ are generators of the chiral
gauge transformations.
${\cal P}$ denotes path ordering, and
$\epsilon$ is a four-vector with infinitesimal components.
We may choose $\epsilon_0 > 0$ so that (\ref{ps}) is in fact
time ordered.
Note that $\left[Q,T^a\right] = 0$, so that the position of
$Q$ in (\ref{ps}) is irrelevent.
Henceforth, for notational simplicity, we shall write the path ordered
exponential
in (\ref{ps}) more compactly as $\PA$.

The counterterm denoted by $\{ c.t.\}$ in (\ref{JJ1})
is the infinite
subtraction which must be made to relate the renormalized operator
$J^{\mu} (x)$ to the infinite operator
$\lim_{\epsilon \rightarrow 0} J^\mu (x\vert \epsilon )$.
As an alternative to equation (\ref{JJ1}),
we can consider defining
\beq
\label{JJ2}
J^{\mu}~\equiv~\lim_{\epsilon\rightarrow 0}
\,\left[\, J^\mu\left(x\,\vert\,\epsilon\right) ~+~
\left\{ c.t. \right\}_{\epsilon} \,\right] ~,
\eeq
where now the counterterm is a function of $\epsilon$, becoming infinite as
$\epsilon \rightarrow 0$.
The counterterm can be chosen so as to cancel the c-number which arises
from normal ordering $J^\mu\left( x\,\vert\,\epsilon\right)$
in a {\it trivial} background $\A \equiv 0$, $\Phi(x) \equiv 0$.

The ABJ anomaly arises from the fact that the c-number
relating $J^\mu\left( x\,\vert\,\epsilon\right)$
and its normal ordered equivalent is dependent on the
background fields $\BF$.
This dependence arises when we choose to expand the field operators
$\Psi\left(x \pm \epsilon/2\right)$
appearing in (\ref{ps}) in terms of mode functions
which are solutions to the Dirac equation in the $\BF$ background.
Thus, our definition of normal ordering depends on the background.
More explicitly, we write the fermionic field operators as a
sum of modes (discrete for finite $L$)
\beq
\label{inmodes}
\Psi(x) ~=~ \sum_n\,
\hat{a}^{\,\rm in}_n\,\Psi^{\,{\rm in},+}_n ~+~
\hat{b}^{\,{\rm in}\dagger}_n\,\Psi^{\,{\rm in},-}_n ~,
\eeq
or
\beq
\label{outmodes}
\Psi(x) ~=~ \sum_n\,
\hat{a}^{\rm out}_n\,\Psi^{{\rm out},+}_n ~+~
\hat{b}^{{\rm out} \dagger}_n\,\Psi^{{\rm out},-}_n ~.
\eeq
The operators $\hat{a}^{\rm in,\, out}_n$ annihilate
in-- or out--fermions and
the operators $\hat{b}^{\,{\rm in,\, out} \dagger}_n$ create in--
or out--anti--fermions, respectively.
These operators define in-- and out--vacua through the
requirements that all annihilation operators annihilate their
respective vacua
\beqa
\label{vacua}
\hat{a}^{\rm\, in}_n\,\vert\, 0\,\rangle^{\rm\, in} ~=~ 0
&{\rm and}&
\hat{b}^{\rm\, in}_n\,\vert\, 0\,\rangle^{\rm\, in} ~=~ 0  ~,\\
\hat{a}^{\rm\, out}_n\,\vert\, 0\,\rangle^{\rm\, out} ~=~ 0
&{\rm and}&
\hat{b}^{\rm\, out}_n\,\vert\, 0\,\rangle^{\rm\, out} ~=~ 0  \nonumber ~.
\eeqa

The spinor functions $\Psi^{\,{\rm in},\pm}$ and $\Psi^{{\rm out},\pm}$
are orthonormal sets of functions which satisfy the Dirac
equation
\beq
\label{Dirac}
\left(\,
\partial\hspace{-2.5mm}/ ~-~  i\, g\, A\hspace{-2.5mm}/ ~+~
\Phi\,\lambda\, P_R ~+~ \lambda^\dagger\,\Phi^\dagger P_L
\,\right)\,\Psi^{{\rm\, in,\, out},\pm}_n(x) ~=~ 0 ~.
\eeq
Here $P_{L,R}$ projects onto left-- or right--handed fermions,
and $\lambda$ is an isospin matrix of Yukawa couplings.
If the background fields approach a vacuum at large $\vert\, t\,\vert $,
the solutions to the Dirac equation can be chosen to
satisfy the boundary conditions
\beqa
\label{TBC}
\Psi^{\,{\rm in},\pm}_n(x) &\rightarrow&
R_{\,\rm in}(\vec{x})~\psi^{\pm}_n(\vec{x})~ e^{\mp iE_nt}
\hspace{0.6cm}{\rm as}\hspace{0.5cm} t\rightarrow -\infty ~, \\
\Psi^{{\rm out},\pm}_n(x) &\rightarrow&
R_{\rm out}(\vec{x})~\psi^{\pm}_n(\vec{x})~ e^{\mp iE_nt}
\hspace{0.5cm}{\rm as}\hspace{0.5cm}t\rightarrow +\infty \nonumber ~.
\eeqa
Here $\psi^{\pm}\left(\vec{x}\right)$ are eigenfunctions of
the free Dirac Hamiltonian with positive  or negative energy
eigenvalues $\pm E_n$.
These functions $\psi^{\pm}\left(\vec{x}\right)$ are complete
and we choose them to be orthonormal functions of $\vec{x}$.
The necessity of the gauge matrix ``twists'' $R\left(\vec{x}\right)$
is central to the phenomena of fermion number violation.
Smooth backgrounds which allow the implementation of free boundary
conditions ($R_{\,\rm in} = R_{\rm out} = {\cal I}$)
at {\it both} early and late times can be seen
to conserve fermion number~\cite{us}.

Substituting (\ref{inmodes}) and (\ref{outmodes}) into (\ref{ps}) results
in an operator which is not normal ordered: it contains $\hat{b}$
operators to the left of $\hat{b}^{\dagger}$ operators.
Normal ordering the operator is equivalent to subtracting a c-number equal
to the infinite sum
\beq
\label{NOsum}
S^{\mu}_{\epsilon} \left[A, \Phi\right] ~\equiv~
\sum_n \, \overline{\Psi}^{\, -}_n (x + \epsilon/2)\,\gamma^\mu\, Q\,\PA\,
\Psi^{-}_n (x - \epsilon/2) ~,
\eeq
where the sum is over either in-- or out--modes depending on whether
the operator is being normal ordered with respect to the $\ivr$ or
$\ovr$ vacuum. Thus,
\beq
\label{NOJ}
\JPS ~=~ :\!\JPS\! :_{\,{\rm in},\,{\rm out}} ~+~
\SA \left[A, \Phi \right]_{\,{\rm in},\,{\rm out}} ~.
\eeq

As mentioned previously,
the background dependence in $\SA\left[ A,\Phi\right]$ comes from the
implicit dependence of the solutions $\Psi^{\pm}$ on the background fields.
In defining the regularized current via (\ref{JJ1}), we must choose a
{\it single} value for the counterterm.
We take this to be the value of (\ref{NOsum}) when the background fields
are trivial ($\A, \Phi (x) = 0$):
\beq
\label{JJS}
J^{\mu} (x) ~=~\lim_{\epsilon \rightarrow 0}~
\left[\,  :\!\JPS\! :_{\,{\rm in},\,{\rm out}} ~+~
\SA \left[ A,\Phi \right]_{\,{\rm in},\,{\rm out}} ~-~
S_{\epsilon}^{\mu} \left[\, 0\,\right]_{\,{\rm in},\,{\rm out}}\,\right]  ~.
\eeq
The anomalous divergence in (\ref{anom}) arises from the finite
dependence of $J^{\mu}(x)$ on $\BF$ which is uncancelled by the
counterterm.

The decomposition (\ref{NOJ}) is gauge invariant in the following sense.
If under a gauge transformation
\beqa
\label{AGT}
A_{\mu}  &\rightarrow & A'_{\mu} (x) ~=~
\Omega\,\A\,\Omega^\dagger ~-~
\frac{i}{g}\,\partial_\mu\Omega\,\Omega^\dagger ~,   \\
\Phi &\rightarrow & \Phi' (x) ~=~ \Omega\,\Phi (x) \nonumber ~,
\eeqa
the fermion modes are also transformed
\beq
\label{PSIGT}
\Psi^{\pm}_n ~\rightarrow~ \Psi^{' \pm}_n ~=~ \Omega~ \Psi^{\pm}_n ~,
\eeq
then each term on the right hand side of (\ref{NOJ}) remains
individually gauge invariant.
In particular, we define operators which count in-- or out--modes
\beq
\label{counter}
:\! \Cop \!:_{\,\rm in,\, out} ~\equiv~
\lim_{\epsilon\rightarrow 0}\int d^3x ~:\!
J^0 \left(x\,\vert\,\epsilon\right)\! :_{\rm\, in,\, out}~.
\eeq
These counter operators are invariant under (\ref{AGT}) and (\ref{PSIGT}),
although the modes that they count are changed.
This definition is quite natural since if the $\Psi^{\pm}_n$ are
initially chosen to be solutions of the Dirac
equation in the $\A$ background they will continue to be solutions
of the Dirac equation in the new gauge background $A'_{\mu} (x), \Phi' (x)$,
now with gauge rotated boundary conditions.

Our general strategy for evaluating the in-- and out--counters
$:\!\Cop\! :_{\,\rm in,\, out}$ will be to find a gauge in which the c-numbers
$\SA \left[ A,\Phi \right]_{\,\rm in,\, out}$ are easily computed.
This will typically be a gauge in which the backgrounds falls
off sufficiently rapidly to allow the use of standard perturbation theory as
we will discuss below.

In the original works of Adler, Bell and Jackiw~\cite{ABJ},
the anomalous divergence resulting from the sum
(\ref{NOsum}) is computed within perturbation theory.
In section 3,  we will show how the sum over exact Dirac solutions can
be expressed perturbatively,
making contact with the earlier approaches.
However, for our purposes it is crucial that the normal ordered part of
the current be retained.
In the vacuum matrix elements studied by \cite{ABJ},
this part of the current is always null.
However, we are interested in amplitudes involving fermion excitations
in the initial and/or final states, and the normal ordered part of the
integrated charge operator $\int d^3 x ~J^0(x)$ provides a counter
$:\!\Cop\! :_{\,\rm in,\, out}$ for these states.

We wish to specify our gauge field background in a spacetime region
like the one depicted in figure 1.
In order to discretize the fermionic energy levels,
we have taken the region to have finite spatial extent $L$,
and imposed periodic boundary conditions on the background and
the solutions $\psi^{\pm} (\vec{x})$.
We will require that {\it all} of $\vec{\Sigma}$,
the timelike boundary of the spacetime region in figure 1,
lie {\it outside} the forward light-cone of regions where $\A$ has its
support.
This will allow us to ignore essentially all activity, either fermionic
or bosonic, on this boundary, simply by invoking causality.
One can regard this condition as a restriction on the support of $\BF$
and/or as a prescription for the order of limits when the timelike and
spacelike extents, $T_f - T_i$ and $L$, of the region are taken to
infinity.
The spatial extent $L$  should be taken to infinity first,
with $T_i$ and $T_f$ fixed, as reflected in figure 1.
Finally, $T_f$ and  $T_i$ are taken to $\pm\infty$, respectively.

\ifx\pansw\pictures
$$
\beginpicture
\setplotarea x from -65 to 65 , y from -40 to 55
\ellipticalarc axes ratio 3:1 360 degrees from 57 100 center at 0 100
\ellipticalarc axes ratio 3:1 -180 degrees from 80 70 center at 0 70
\ellipticalarc axes ratio 3:1  50 degrees from 80 70 center at 0 70
\ellipticalarc axes ratio 3:1 -50 degrees from -80 70 center at 0 70
\ellipticalarc axes ratio 3:1 180 degrees from -80 10 center at 0 10
\ellipticalarc axes ratio 3:1 180 degrees from -57 -20 center at 0 -20
\setdots
\ellipticalarc axes ratio 3:1 180 degrees from 80 70 center at 0 70
\ellipticalarc axes ratio 3:1 180 degrees from 80 10 center at 0 10
\ellipticalarc axes ratio 3:1 180 degrees from 57 -20 center at 0 -20
\setsolid
\putrule from 80 10 to 80 70    \putrule from -80 10 to -80 70
\arrow < 4 pt> [0.4,1.5] from 0 120 to 0 140  \put{$t$} at 0 150
\arrow < 4 pt> [0.4,1.5] from 100 40 to 120 40 \put{$r$} at 130 40
\arrow < 4 pt> [0,0] from 0 40 to 80 120
\arrow < 4 pt> [0,0] from 0 40 to 80 -40
\arrow < 4 pt> [0,0] from 0 40 to -80 120
\arrow < 4 pt> [0,0] from 0 40 to -80 -40
\putrule from -35 70 to -28 70   \putrule from 28 70 to 35 70
\putrule from -35 10 to -28 10     \putrule from 28 10 to 35 10
\putrule from 90 70 to 100 70   \put{$T_f$} at 110 70
\putrule from 90 10 to 100 10    \put{$T_i$} at 110 10
\put{$\vec{\Sigma}$} at -100 40
\put{ Figure 1: Spacetime region with spatial boundary $\vec{\Sigma}$.}
at 0 -80
\put{The ratio of the radii of cylinder and cone reflects our order of
limits.} at 20 -100
\endpicture
$$
\bigskip
\else\fi

\mysection{Perturbation Theory and Selection Rule}

The central result of this section will be a selection rule
for fermionic scattering amplitudes in backgrounds
$\BF$ which are solutions to the classical field equations.
The results can also be generalized to  a larger class of configurations
which are not necessarily solutions, as long as they have the
appropriate asymptotic behavior.
As we will discuss in section 4,
classical solutions which in the unitary gauge asymptotically
approach the unique unitary gauge vacuum
\beqa
\label{Uvac}
\A &=& 0 ~, \\
\Phi (x) &=& \langle\,\Phi(x)\,\rangle ~=~ v~{\cal I} \nonumber ~,
\eeqa
will do so
sufficiently rapidly that the solutions to the
Dirac equation are compatible with free boundary conditions
as in (\ref{TBC}), but with $R(\vec{x}) = {\cal I}$.
Any configuration which can be placed in unitary gauge in all of
spacetime via a smooth gauge transformation
therefore will not lead to fermion number violation \cite{us}.
However, configurations with nontrivial topology\footnote{Any gauge
configuration which approaches
a pure gauge given by a gauge function $\Omega(x)$
as $|x| \rightarrow \infty$, such that the gauge function
$\Omega (x)$ provides a map from $S^3 \rightarrow SU(2)$, can
be classified topologically.}
cannot be globally transformed to a gauge with a unique vacuum
without a discontinuity in the gauge transform.
(See \cite{JMR} for an explicit example
of how this works in pure Yang--Mills theory.)
Such topologically nontrivial configurations, which in $A_0 = 0$
gauge interpolate between different $n$ vacua, are known
to lead to fermion number violation \cite{us,NC}.

Hence, we will be interested in configurations which can be
transformed to unitary gauge {\it either} at early or late
times, but not both simultaneously.
For simplicity, we will take the configurations to be in
unitary gauge initially,
so that they approach (\ref{Uvac}) at early times
($t \rightarrow - \infty$).
At late times,
the configuration will approach a gauge transform
of the vacuum (\ref{Uvac}):
\beqa
\label{fvac}
\A &\rightarrow&  - \frac{i}{g}\,\Omega (\vec{x})\,
\partial_\mu \Omega (\vec{x})^\dagger  ~, \\
\Phi (x) &\rightarrow&  v~\Omega (\vec{x}) ~{\cal I} \nonumber ~.
\eeqa
Any nontrivial topology will be reflected in the winding number of the
gauge function $\Omega (\vec{x})$. Solutions to the Dirac equation
in this class of backgrounds will satisfy boundary conditions given
by (\ref{TBC}) with $R_{\rm in} (\vec{x}) = 1$ and
$R_{\rm out} (\vec{x})= \Omega (\vec{x})$.

Let us define in-- and out--fermion states by application of the
Heisenberg operators defined previously in (\ref{inmodes}) and
(\ref{outmodes}):
\beq
\label{state1}
\vert\, i\, \rangle^{\rm\, in} ~=~
\prod_{l}^{N_i} ~\prod_{m}^{\bar{N}_i} ~
\hat{a}^{\,{\rm in} \dagger}_{n_l} ~
\hat{b}^{\,{\rm in} \dagger}_{\bar{n}_m} ~  \ivr
\eeq
and
\beq
\label{state2}
^{\rm out} \langle\, f\,\vert ~=~  \ovl  ~
\prod_{l}^{N_f} ~\prod_{m}^{\bar{N}_f}
   ~\hat{a}^{\,\rm out}_{n_l}  ~ \hat{b}^{\,\rm out}_{\bar{n}_m} ~
\eeq
For every $l$ ($m$),
the label $n_l$ ($\bar{n}_m$) represents a set of numbers specifying
a one fermion (one anti-fermion) state.
The states (\ref{state1}) and (\ref{state2}) are clearly eigenstates
of the operators which count in-- and out--modes,
$\hat{a}^{\,{\rm in}\dagger}\,\hat{a}^{\,{\rm in}}$ and
$\hat{b}^{\,{\rm in}\dagger}\,\hat{b}^{\,{\rm in}}$,
and
$\hat{a}^{\,{\rm out}\dagger}\,\hat{a}^{\,{\rm out}}$ and
$\hat{b}^{\,{\rm out}\dagger}\,\hat{b}^{\,{\rm out}}$, respectively.
The relation of these operators to the fermion number charge will be
explicated below.
More complicated states which are not eigenstates of these operators
may be formed as superpositions of the states given above.

Our goal is to compute the matrix element given by
\beq
\label{SR}
^{\rm out} \langle\, f\,\vert
\,\left(\,\hat{C}_f ~-~\hat{C}_i\,\right)\,
\vert\, i\,\rangle^{\rm\, in},
\eeq
where $\hat{C}_{f,i} \equiv \int d^3x~ J^0 (\vec{x}, T_{f,i}  )$.
We choose to write $J^0 (\vec{x}, T_{f}  )$ in terms of an operator
which is normal ordered with respect to $\ovr$,
while $J^0 (\vec{x}, T_{i}  )$ is written in terms of an operator
which is normal ordered with respect to $\ivr$.
Using the definition of $J^0 (x)$ given by (\ref{JJS}),
and the orthogonality
of mode functions $\Psi^{\,{\rm in},\pm}$ and
$\Psi^{{\rm out},\pm}$, we have
\beqa
\label{SR1}
\lefteqn{^{\rm out} \langle\, f\,\vert  \,
\left(\,\hat{C}_f ~-~\hat{C}_i\,\right)\,
\vert\, i\,\rangle^{\rm\, in} ~=} & & \\
& &  S_{fi} ~  \left\{\,
Q_f ~-~ Q_i  ~+~  \lim_{\epsilon \rightarrow 0}
\int d^3x ~  \left(\,
S^0_{\epsilon} \left[A,\Phi(\vec{x},T_f)\right]_{\,\rm out} ~-~
S^0_{\epsilon} \left[A,\Phi(\vec{x},T_i)\right]_{\,\rm in}
\,\right) \,\right\} ~. \nonumber
\eeqa
Here $Q_{f,i}$ are the eigenvalues of the
normal ordered part of the counter
operator:
\beq
\label{CQ}
:\!\Cop_{f,i}\! :\,\vert\, f,i\,\rangle^{{\rm out},\,{\rm in}} ~=~
Q_{f,i}\,\vert\, f,i\,\rangle^{{\rm out},\,{\rm in}},
\eeq
and $S_{fi} ~=~ ^{\rm out}\langle\, f\,\vert\, i\,\rangle ^{\rm\, in}$
is the S-matrix amplitude between the chosen in- and out- states.
For the in-- and out-- states we have defined
$Q_{f,i} \propto\left( N_{f,i} - \bar{N}_{f,i}\right)$,
with $Q_{f,i}$ taking on integer values when the charge Q
appearing in (\ref{ps}) is an integer~\footnote{We will assume for
simplicity that this is the case.  It can be easily generalized.}.

To make further progress, we must compute the difference of
the c-number functions $\SA \left[A,\Phi \right]$ which appear
in the integral in (\ref{SR1}). As will be discussed at the
end of this section,
$\SA \left[A,\Phi \right]$ can be computed directly in
perturbation theory when $\BF$ approach their
unitary gauge vacuum values rapidly enough.
The result will be
\beq
\label{pt}
\SA \left[A,\Phi \right] ~=~ N\, K^{\mu} ( A ) ~+~ L^{\mu} (A,\Phi) ~,
\eeq
where $K^{\mu}$ is the usual topological current density
($\partial\cdot K \equiv {g^2\over 16\pi^2}\,\ffdool$) and $L^\mu$ is
divergenceless ($\partial\!\cdot\! L  = 0$).  This result is
obtained whether the normal ordering is with respect to the in--
or the out--vacuum, as we will explain below.

This result can be directly applied to
$S^0_{\epsilon} \left[A,\Phi(\vec{x},T_i)\right]_{\,\rm in}$.
However,
the background fields at late times approach a gauge transform
of the unitary gauge vacuum (see (\ref{fvac})), so we
must use the following trick to evaluate
$S^0_{\epsilon} \left[A,\Phi(\vec{x},T_f)\right]_{\,\rm out}$.
We use the fact that $\SA \left[A,\Phi \right]$ is gauge invariant
(see (\ref{NOsum})). Therefore,
\beq
\label{gtwist}
S^0_{\epsilon} \left[A,\Phi(\vec{x},T_f)\right]_{\,\rm out} ~=~
S^0_{\epsilon} \left[A',\Phi'(\vec{x},T_f)\right]_{\,\rm out} ~.
\eeq
The primed fields are obtained by applying the inverse of the
gauge transform function $\Omega(\vec{x})$ defined in (\ref{fvac}):
\beqa
\label{primebf}
A'_\mu (\vec{x},T_f) &=& \Omega^\dagger\, A_\mu (\vec{x},T_f)\,\Omega ~-~
\frac{i}{g}\,\partial_i\,\Omega^\dagger\,\Omega  \\
\Phi' (\vec{x},T_f) &=& \Omega^{\dagger}\,\Phi (\vec{x},T_f) \nonumber ~.
\eeqa
The primed fields thus approach the unitary gauge
vacuum as $t \rightarrow \infty$, and the result
(\ref{pt}) can be applied with the RHS now a function of
the primed fields $A',\Phi'$.

The integral in (\ref{SR1}) thus becomes
\beq
\label{SRintegral}
\label{diff}
 q (A'_f) ~-~ q (A_i) ~,
\eeq
where the Chern--Simons number $q$ is defined as
\beq
\label{q}
q\left(A\right) ~\equiv~
\int d^3x~ K^0 ~=~
{g^2\over 16\pi^2}\int d^3x~\epsilon^{0ijk}~
{\rm Tr}\,\left(A_iF_{jk} ~+~ \frac{2}{3}\, i\, g\, A_iA_jA_k\right).
\eeq
To arrive at this result we used the fact that any integrals over the
surface $\vec{\Sigma}$ vanish by causality and that the primed fields
$\{ A',\Phi' \}$ at $T_f$, can be smoothly connected to
the fields $\{ A,\Phi \}$ at $T_i$. This fact and
the vanishing divergence of $L^\mu$ are sufficient to guarantee the
cancellation of the integrals $\int d^3x~ L^0$ at early and late times.

To simplify (\ref{SRintegral}) further, we use the following result:
\beq
\label{qtransf}
q\left(A'\right) ~=~
q\left(A\right) ~+~ \nu\left(\Omega^\dagger\right) ~+~
{i\,g\over 8\pi^2}\int_S d\sigma_i~ \epsilon_{ijk}~{\rm Tr}~
\Omega\partial_j\Omega^\dagger~A_k
\eeq
where the winding number of the map $\Omega^\dagger$ is
\beq
\label{wind}
\nu\left(\Omega^\dagger\right) ~=~ - \nu\left(\Omega\right) ~=~
- \int {d^3x\over 24\pi^2}~ \epsilon_{ijk}~{\rm Tr}~
\Omega^\dagger\partial_i \Omega~\Omega^\dagger\partial_j \Omega~
\Omega^\dagger\partial_k \Omega ~.
\eeq
When the map $\Omega$ is constant on the $2$--sphere $S$ at spatial
infinity,
the surface term in (\ref{qtransf}) vanishes.
Further, (\ref{wind}) is then integer-valued, according to homotopy arguments.
So we have $\nu\left(\Omega\right) ~=~  n$.
{}From our definition of the vacua in (\ref{Uvac}) and (\ref{fvac}),
one can see that
$\nu\left(\Omega\right)$ is precisely
the Higgs winding number of the configuration $\BF$.
This agrees with the recent result of \cite{5authors}.

Thus, (\ref{SRintegral}) can be rewritten as
\beq
\label{diff1}
 q (A_f) ~-~ q (A_i) ~-~ n.
\eeq
We can also apply the operator relation (\ref{anom}) directly to the matrix
element (\ref{SR}). This yields~\footnote{Our calculation is
performed in a fixed volume as defined in section 2, so
four--integrals like the topological charge are well-defined
at each step. Since the topological charge eventually
cancels in our selection rule (\ref{SR4}),
we need not be concerned with whether the integrals in
(\ref{ffqq}) continue to be well-defined when we take the
volume to infinity.
For a discussion of this possibility, see \cite{5authors}.}
\beq
\label{SR2}
 ^{\rm out} \langle\, f\,\vert  \int d^4x~ \partial_0 J^0 (x)
\,\vert\, i\,\rangle^{\rm\, in}
{}~=~  \left(\, \frac{g^2 N}{16\pi^2}
\int d^4x~\ffdool \,\right) ~S_{fi} ~+~
^{\rm out} \langle\, f\,\vert\,
\int d \vec{\Sigma} \cdot \vec{J}\,\vert\, i\,\rangle^{\rm\, in} ~,
\eeq
where again the last integral over $\vec{\Sigma}$ is zero.
Combining the results (\ref{SR1}), (\ref{diff1}) and (\ref{SR2})
yields
\beq
\label{SR3}
\left(\, Q_f ~-~ Q_i\,\right)~S_{fi} ~=~
\left[\, n\, N ~+~
\frac{g^2 N}{16 \pi^2}\,\int d^3x~\ffdool
{}~-~ N\,\left(\, q (A_f) ~-~ q (A_i) \,\right) \,\right] ~S_{fi} ~.
\eeq
Since $\A$ has no support on the spatial boundary $\vec{\Sigma}$,
we have
\beq
\label{ffqq}
\frac{g^2}{16 \pi^2}\int d^4x~ \ffdool ~=~ q (A_f) -q (A_i) ~.
\eeq
Our final selection rule is thus
\beq
\label{SR4}
\left(\, Q_f ~-~ Q_i\,\right)~S_{fi} ~=~ n\, N\, S_{fi} ~.
\eeq
As expected, the amount of fermion number violation is determined
by the change in winding of the two vacuums (\ref{Uvac}) and
(\ref{fvac}). We note that the selection rule (\ref{SR4}) implies
an index theorem for the operator $\left( \partial_t - H_t \right)$,
where $H_t$ is the time-dependent
Hamiltonian of the fermion field, as discussed in \cite{NC}.
The net production of fermions is equal to the spectral flow
of $H_t$.

In the remainder of this section,  we will show how the c-number functions
$\SA [A]$ can be computed in perturbation theory, culminating in the
expression (\ref{pt}).
We first express $\SA [A]$ directly in terms of vacuum matrix elements:
\beq
\label{vev}
\SA\left[A\right]_{\rm\, in,\, out} ~=~ \ovl ~
\overline{\Psi} (x + \epsilon/2)\,\gamma^\mu Q \,{\cal P}(A)
\,\Psi (x - \epsilon/2)~\ivr ~.
\eeq
It is clear that the normal ordered part of the current
$:\!\JPS\! :$
is annihilated when it acts either to the right
if we normal order with respect to in--states, or
to the left if we normal order with respect to out--states.
Thus,
what remains in the above matrix element is just
$\SA[A]_{\,\rm in}$, which therefore is identical to
$\SA[A]_{\,\rm out}$.

To compute the matrix element in (\ref{vev}) using
perturbation theory,
we first go to the Interaction picture.
Until now our discussion has been in the Heisenberg picture,
with field operators evolving in time as
$\Psi(\vec{x},t) = U (t,t_0)\,\Psi(\vec{x},t_0)\, U^\dagger(t,t_0)$,
where the time evolution operator for the time dependent Hamiltonian
is $U(t',t)\equiv \exp\, i\int^{t'}_t dt'' H(t'') $.
The vacuum and all other states are time independent.
We denote fields in the Interaction picture by $\phi(t)$, with
\beqa
\label{Ipic}
\phi(\vec{x},t)^{\rm\, in,\, out} ~=~
U_I(t,\pm\infty)~\Psi(\vec{x},t)^{\rm\, in,\, out}~
U_I^{\dagger}(t,\pm\infty) ~.
\eeqa
Here $H_0$ is the free Hamiltonian, and the evolution operator
$U_I(t',t) ~=~ \exp{-i\int^{t'}_t dt'' H_I(t'')}$ is given in terms of the
interacting part of the Hamiltonian.
The (time dependent) Interaction picture vacuum is given by
\beq
\label{Ivac}
\vert\, 0,t\,\rangle ~=~ U_I (t, -\infty)\,\ivr ~=~ U_I (t,+\infty)\,\ovr.
\eeq

Substituting (\ref{Ipic}) into the matrix element (\ref{vev}) yields
\beqa
\label{vev1}
\lefteqn{\SA \left[ A \right] ~=} & & \\
& & \langle\, 0,t\,\vert\,  U_I^{\dagger} (-\infty,+\infty)~
T\left\{\bar{\phi}(x+\epsilon/2)\,\gamma^{\mu}
Q\, {\cal P}(A)\,\phi (x-\epsilon/2)\, U_I(t,-t)\right\}\,
\vert\, 0,-t\,\rangle \nonumber ~.
\eeqa
Expanding the exponential $U_I(t,-t)$
inside the time ordered product reproduces ordinary perturbation
theory in terms of free Feynman propagators and
insertions of the background $\A$.
The factor of $U_I^{\dagger} (-\infty,+\infty)$
divides out disconnected graphs in the usual way.

One can now explicitly evaluate (\ref{vev1}) using standard Feynman rules.
To derive (\ref{pt}), we simply note that the
graphs which correspond to the computation of the {\it divergence}
of (\ref{pt}) $\partial_\mu\, S^\mu_\epsilon$
are precisely those computed in the usual perturbative derivation
of the anomaly~\cite{ABJ}.

\mysection{Asymptotics of Unitary Gauge Solutions}

In this section,
we study the asymptotic behavior of Minkowski dissipative solutions
to the Yang--Mills--Higgs equations in the unitary gauge.
Our purpose is to check that they dissipate rapidly enough to allow
the use of ordinary perturbation theory as applied in the previous section.
The issue here is the use of the Interaction picture, which
assumes that the Heisenberg field $\Psi (x)$ approaches
the free field $\phi (x)$ at early and late times.
We need to check that the
background fields approach their unitary gauge vacuum values
rapidly enough that solutions to the Dirac equation (\ref{Dirac})
are compatible with free boundary conditions like those in (\ref{TBC}),
but with $R(\vec{x}) = {\cal I}$.

To study the behavior of the $U$-matrix in an arbitrary background, we will
use the Yang--Feldman equation which relates the Heisenberg field to a
perturbative expansion in terms of the free field boundary condition, the free
propagator, and the background $\BF$. We write
\beqa
\label{YF}
\lefteqn{\Psi_n (\vec{x},x_0 ) ~=} & & \\
& & \phi_n (\vec{x},x_0) ~+~ \int d^4 y~
\Delta (x-y)\,\left(\, g\, A_{\mu} (y)\, \gamma^{\mu} P_L ~+~
\Phi\,\lambda\, P_R ~+~ \lambda^\dagger\,\Phi^\dagger P_L
\,\right)\,\Psi_n (y) \nonumber ~.
\eeqa
It is straightforward to check that $\Psi_n (x)$
defined in this way satisfies the
Dirac equation in the $\BF$ background.
The choice of free propagator $\Delta (x-y)$
determines the boundary conditions that $\psi_n (x)$ satisfies.
For an advanced propagator,
with support only for $y_0 > x_0$, (\ref{YF}) implements
the advanced boundary condition that
$\Psi_n (\vec{x},x_0) \rightarrow \phi_n (\vec{x},x_0)$
as $x_0 \rightarrow\infty$,
provided that the integral on the RHS of the equation approaches
zero in the same limit.
If the integral does not vanish in that limit,
the Dirac equation in the $\BF$ background is incompatible with free
boundary conditions in the far future.
Similar considerations apply for the retarded propagator and
retarded boundary conditions.
The integral in (\ref{YF}) can be expanded perturbatively
by using (\ref{YF}) recursively. This yields an infinite sum
of multiple integrals, each of which must vanish in the limit
that $x_0 \rightarrow \infty$.

In the present case, the propagator $\Delta (x-y)$
can be either that of a
massive or massless fermion since the spontaneously
broken theory can accomodate both.
For example, in the minimal electroweak theory,
the neutrino component of the lepton doublet
is massless within a perturbative expansion about the vacuum,
while its partner is massive.
The Yang--Feldman equation for a massless fermion was studied
previously in \cite{us}.

In unitary gauge,  the linearized equations of motion
reduce to the free massive wave equation for each degree of freedom.
Therefore, at very early or late times, we can write any disspative
solution in the form
\beq
\label{UGSol}
A ( x ) ~=~ \int d^3 k~  A (\vec{k}) ~
e^{i\, k\cdot x} ~+~ \{ {\rm h.c.} \} ~,
\eeq
where the four vector $k_{\mu} = ( \omega_k, \vec{k} )$
and $\omega_k^2 = M^2 + \vec{k}^2$.
We intend (\ref{UGSol}) to imply similar expressions for
each of the fields $\BF$.
Here $A$ represents any of the background fields since they
each obey the same linearized equation in unitary gauge.
We will suppress gauge and Lorentz indices for simplicity.
Finally, we will assume that the momentum space function $A (\vec{k})$
is smooth and integrable, with the same being true of
its first few derivatives.

Now consider the large $x_0$ behavior of the above integral.
Let us consider the behavior of the gauge potential along
rays given by $\vert\,\vec{x}\,\vert = a x_0$, where $a$ is some constant.
Rewriting (\ref{UGSol}) in spherical coordinates yields
\beq
\label{UGS1}
A (\vec{x}, x_0) ~=~ \int dk~ d \phi~ \int_{-1}^{+1} du ~k^2~
A (k, \phi, u) ~ e^{i\, k\, x_0\, (\, au\, -\, 1\, )}
{}~+~ \{ {\rm h.c.} \} ~.
\eeq
One can compute the asymptotic behavior of the integral
(\ref{UGS1}) by making the stationary phase approximation
on the argument of the exponential as $x_0 \rightarrow \infty$.
We find that $A (\vec{x}, x_0)$ has support within the
forward lightcone ($a < 1$) which falls off uniformly
at least as fast as $1/ |x_0|^{3/2}$.
Outside the lightcone, the function falls off exponentially.

The analysis of massless propagators in the expansion of (\ref{YF}) has
been performed in \cite{us}. The asymptotic behavior derived above
satisfies the conditions derived previously
for that case, so we will concentrate
on the case of masssive propagators $\Delta (x-y)$ which
have the following behavior inside the lightcone
\beq
\label{mprop}
\Delta (y-x) ~=~ \frac{\theta (y_0 - x_0)}{4 \pi r}\,
\frac{\partial}{\partial r}\, J_0\left(\, m \sqrt{t^2 -r^2}\,\right) ~,
\eeq
where $r = \vert\,\vec{y} - \vec{x}\,\vert $, $t = y_0 - x_0$ and
$J_0 (x)$ is a Bessel function.
Deep inside the lightcone ($t >> r$) the propagator falls off
like $1/t^{3/2}$, while near the lightcone ($t \sim r$) it
falls off like $1/t$.
Outside the lightcone $\Delta (y-x)$ is zero.

Terms in the expansion of (\ref{YF})
are composed of nested integrals are of the form
\beqa
\label{YFexp}
\lefteqn{\int d^4 y_1~ \Delta (y_1 - x)\, A(y_1)\,\int d^4 y_2 ~
\Delta (y_2 - y_1)\, A(y_1) \cdots}  & & \\
& & \hspace{5cm}
\int d^4 y_{n-1}~ \Delta (y_n - y_{n-1})\,
A(y_n)\,\phi_n (y_n) \nonumber ~.
\eeqa

For the case of massive propagators and the asymptotic
behavior of the background fields derived above,
the lowest order term of the type (\ref{YFexp})
takes the form
\beq
\label{YFI1}
\int_{x_0}^{\infty} dy_0 ~\frac{1}{y_0^{3/2}}~
\int_R d^3 \vec{y}~~
\frac{1}{r}\frac{\partial}{\partial r}
\left(\, J_0\left(\, m \sqrt{t^2 -r^2}\,\right)\,\right)\,
g\left( \vec{y}, y_0\right)~
e^{i \vec{p}_n\cdot \vec{y}}\, e^{i E_n y_0} ~,
\eeq
where $g\left(\vec{y}, y_0\right)$ is a bounded
function derived from the asymptotic
expansion of (\ref{UGS1}).
Subleading terms in (\ref{YFI1}) have been suppressed.
The region of spatial integration
$R$ is determined by the intersection of the volume of the
forward lightcone
of $x^\mu$:
$\vert\,\vec{y} - \vec{x}\,\vert < \left( y_0 - x_0 \right)$
and the volume within which the background has its
support: $\vert\,\vec{y}\,\vert < y_0 $.
The size of the region $R$ is at most $\vert\, y_0 - x_0\,\vert $.
The spatial integration over the region $R$
can be rewritten in spherical coordinates with
origin at $\vec{x}$,
\beq
\label{spi}
\int_R dr ~r^2~ d\phi~ \int_{-1}^{+1} du~
 \frac{1}{r}\frac{\partial}{\partial r}
\left(\, J_0\left(\, m \sqrt{t^2 -r^2}\,\right)\,\right)~
g\left( r, \phi, u\right)~ e^{i p_n\, r\, u}~
e^{i \vec{p}_n \cdot \vec{x}} ~,
\eeq
where $p_n = \vert\,\vec{p}_n\,\vert$.
By partial integration with respect to $u$ we can introduce
another power of $1/r$ in the integrand. We see that the
spatial integral grows at most like $y_0$.
In that case the remaining integral over $d y_0$ vanishes
in the limit that $x_0 \rightarrow \infty$, behaving as
$\sim~ e^{iE_n\, x_0} / x_0^k$, where $k$ is at least $1/2$.
Subsequent multiple integrations in the nesting (\ref{YFexp})
do not alter this behavior, so each term in the expansion
of (\ref{YF}) falls off at least as fast as $1/x_0^{1/2} $.

Since interactions with the background can change a massless
fermion to a massive one, or vice versa (charge current interactions
in electroweak theory), it is also necessary to study
integrals like those above but with a mixture of massive and
massless propagators. It can be shown that these integrals
also approach zero as $x_0 \rightarrow \infty$.

Thus, it appears that unitary gauge solutions which  approach
the vacuum do so sufficiently rapidly to allow the construction
of a time evolution operator $U(t,t')$ with suitable
boundary conditions and hence the manipulations
used in section 3 to perform perturbative interaction picture
calculations.

\mysection{Conclusions}

In this paper,
we have derived selection rules which describe the violation of
fermion number in Minkowski space, in the presence of background
gauge and Higgs fields as in the electroweak theory.
This represents an extension of our previous results in pure
gauge theories to those with spontaneously broken symmetries~\cite{us}.

We have found that for configurations which approach a vacuum
at large times,
up to  additional ``radiation'' which dissipates sufficiently rapidly,
the amount of fermion number violation is determined solely by the
change in Chern--Simons number of the relevant {\it vacua} that the
configuration approaches at early and late times.
In particular,
our new selection rules apply to any classical
solutions which approach vacua in both the far past and future.
Such classical solutions are of interest because they
provide saddlepoints for the evaluation of scattering amplitudes
in the theory with quantized gauge and scalar bosons \cite{GHP}.

A recent paper by Farhi et al.~\cite{5authors} comes to similar
conclusions although with somewhat different methods of analysis.
In particular,
we note that the methods in \cite{5authors} rely on the existence
of a mass gap in the fermion spectrum, and as such cannot accomodate
an electroweak theory with massless neutrinos.
Our methods do not require this gap and apply equally well to theories
with any spectrum of fermion masses.

\newpage
\centerline{\bf Acknowledgements}
\vskip 0.1in
The authors would like to thank Edward Farhi,
Valya Khoze, Krishna Rajagopal and Bob Singleton for useful discussions.
TMG acknowledges the support of the National Science
Foundation under grants \mbox{NSF-PHY-90-96198} and
\mbox{NSF-PHY-94-04057}.
SDH acknowledges support from the National Science Foundation
under grant \mbox{NSF-PHY-87-14654},
the state of Texas under grant \mbox{TNRLC-RGFY-106} and
the Harvard Society of Fellows.

\nc{\ib}[3]{        {\em ibid. }{\bf #1} (19#2) #3}
\nc{\np}[3]{        {\em Nucl.\ Phys. }{\bf #1} (19#2) #3}
\nc{\pl}[3]{        {\em Phys.\ Lett. }{\bf #1} (19#2) #3}
\nc{\pr}[3]{        {\em Phys.\ Rev.  }{\bf #1} (19#2) #3}
\nc{\prep}[3]{      {\em Phys.\ Rep.  }{\bf #1} (19#2) #3}
\nc{\prl}[3]{       {\em Phys.\ Rev.\ Lett. }{\bf #1} (19#2) #3}


\begin{thebibliography}{99}
\bibitem{us} T.M. Gould and S.D.H. Hsu, {\em Anomalous Violation of
Conservation Laws in Minkowski Space}, HUTP-A0/036, JHU-TIPAC-940017.
\bibitem{NC} N.H. Christ, \pr{D21}{80}{1591}
\bibitem{ABJ} S.L. Adler, \pr{177}{69}{2426};
J.S. Bell and R. Jackiw, {\it Nuovo Cimento} {\bf 60A} (1969) 47;
See also R. Jackiw and K. Johnson, \pr{182}{69}{1459};
S.L. Adler and W.A. Bardeen,  \pr{182}{69}{1517};
W.A. Bardeen, \pr{184}{69}{1848}.
\bibitem{JMR} R. Jackiw, I. Munzinich and C. Rebbi, \pr{D17}{78}{1576}.
\bibitem{5authors}
E. Farhi, J. Goldstone, S. Gutmann, K. Rajagopal and R. Singleton,
{\em Fermion Production in the Background of Minkowski Space
Classical Solutions in Spontaneously Broken Gauge Theory
}, MIT preprint CTP-2370, hep-ph/9410365.
\bibitem{GHP} T.M. Gould, S.D.H. Hsu, and  E.R. Poppitz,
{\em Quantum Scattering from Classical Field Theory},
Harvard preprint HUTP-A0/007, hep-ph/9403353.
\end{thebibliography}
\end{document}